\documentclass[aps,twocolumn,showpacs,amsmath,amssymb,floatfix,superscriptaddress]{revtex4}
\usepackage{epsfig} \usepackage{bm}

\usepackage[normalem]{ulem} %
\usepackage{color}

\IfFileExists{srcltx.sty}{\usepackage[active]{srcltx}}

\begin{document}
\title{Effective field theories for the $\nu = 5/2$ edge.}

\author{Alexey Boyarsky},
\affiliation {Institute of Theoretical Physics, ETH H\"{o}nggerberg, CH-8093 Zurich, Switzerland}
\affiliation{Bogolyubov Institute for Theoretical Physics, Kiev 03680, Ukraine}
\author{Vadim Cheianov}
\affiliation{Physics Department, Lancaster University
Lancaster LA1 4YB, UK
}
\author{J\"{u}rg Fr\"{o}hlich}
\affiliation {Institute of Theoretical Physics, ETH H\"{o}nggerberg, CH-8093 Zurich, Switzerland}

\begin{abstract}
We present a list of possible effective theories which may describe a QHE edge at
filling fraction $\nu =5/2$. We show that there exist several abelian and non-abelian
effective theories (apart from those discussed in the literature)
compatible with the physical requirements imposed by the microscopic nature
of the system. We compare predictions of these theories with previous proposals
and with the results of recent experiments. We identify a set of theories,
both abelian and non-abelian, that cannot be distinguished based on
the quasihole tunneling data only. We discuss what experimental
information may be useful in resolving the ambiguity.
\end{abstract}


\maketitle One of the most striking predictions of the theory of the
fractional quantum Hall effect is that a two-dimensional electron
gas in a strong magnetic field may exhibit a ground state supporting
excitations (quasi-particles) with non-abelian braid statistics. In
particular, Wen~[\onlinecite{WenSU22}] and Moore and Read~ [\onlinecite{Moore-Read}]
have brought forward arguments attempting to explain
the $\nu=5/2$ conductance plateau as a manifestation of such a
``non-abelian`` state. Not only is this
possibility fascinating from the theoretical point of view, it also
holds promise for concrete implementations of topologically protected
quantum algorithms [\onlinecite{QI}].

Among non-abelian quantum Hall states, perhaps the simplest is the
Moore-Read (Pfaffian) state, which
has a simple intuitive interpretation as a p-wave paired state of
composite fermions. The intuitive appeal
of the Moore-Read state does, however, not give a compelling clue as
to its microscopic justification. The Hamiltonian
of the many-electron system in a strong magnetic field is not
amenable to a well-controlled perturbative treatment, whether in the
electron or in the composite-fermion basis. Exact
diagonalization numerical studies do not allow one to make fully
reliable statements about macroscopically large systems. In this
situation the decisive word about the nature of the $\nu=5/2 $ state
must come from experiments. A natural target for experimental
investigations is the edge of an incompressible quantum Hall fluid
supporting gapless excitations, including electric currents, and fractionally charged
quasi-holes.

In a recent experiment [\onlinecite{Radu2008}], properties of the
$\nu=5/2$ state were investigated by means of a transport
measurement in a quantum Hall sample with a narrow constriction. The
parameters of the constriction were tuned in such a way that it
served as a weak link between two $\nu=5/2$ quantum Hall edges. The
electrical conductance of the constriction exhibits a zero bias
peak, whose scaling with temperature is consistent with the
assumption that the current is due to weak tunneling of
fractionally charged quasi-particles. By fitting the shape of the
zero-bias conductance peak to the predictions of a model-independent
theory [\onlinecite{Wen:91}] at five different temperatures the
experimentalists produced a two-parameter confidence map for the
electric charge $e^*$ and the scaling dimension, $g$, of the {\it most
relevant} (in the renormalization group sense) quasi-particle tunneling operator.
Knowledge of these two parameters, within experimental errors,
narrows down the list of candidate effective theories significantly.

In this paper we revisit the theory of the $\nu=5/2$ state
taking the confidence map produced in the experiment \cite{Radu2008}
as a starting point. Rather than relying on prejudices based on aesthetic, microscopic or
numerical arguments to sift out candidate theories, we present a list of
theories satisfying a certain minimal set of physical assumptions
and being in reasonable agreement  with the results of [\onlinecite{Radu2008}].
We discuss our results
in the context of a search for non-abelian states.


{\bf General principles.} We first recall some general principles
underlying the effective field theory of the
Quantum Hall edge (for details see~[\onlinecite{Frol,Wen}]) and then
focus on their application to the $\nu=5/2$ state. We assume
that the effective theory of a QHE edge is a chiral conformal field
theory (CFT) that meets the following requirements imposed by
fundamental properties of the system.
\newline (A)~The CFT at the edge
supports a \emph{chiral} current $J$ that is not
conserved due, to the inflow of a Hall current from the
incompressible bulk.  It is convenient to use a \emph{chiral} Bose
field $\phi$ related to the charge density $J^{0}$ by
\begin{equation}
J^0(x)=
  \frac{\sqrt \nu}{2\pi}\partial_x \phi (x),
  \label{J0}
\end{equation}
where $x$ is a natural parameter along the edge and
$\phi$ satisfies the commutation relations
\begin{equation}
\qquad [\phi(x), \phi(x')]= i \pi \mathrm{sign}(x-x').
\label{comphi}
\end{equation}
The factor $\sqrt{\nu}$ in~(\ref{J0}) is dictated by the electric charge conservation
(anomaly cancellation) in the system.

\noindent (B) Since microscopically the system is composed of electrons, the chiral CFT
must contain a local operator $\psi_e$ of unit charge representing the electron in the effective theory
\begin{equation}
[J_0(x), \psi_e(y)]=-\delta(x-y) \psi_e(y). \label{charge1}
\end{equation}

\noindent (C) Fundamentally, any correlation function of the theory
containing an electron-operator insertion must be a single-valued function of the position of
the insertion. In the effective theory language this
means that $\psi_e$ must be \emph{local} with respect to all primary fields of the CFT.

A physically plausible effective theory should satisfy certain
minimality conditions. Complicated theories with very rich
spectra of quasi-partilces, large  central charge and large scaling dimension of electron operators may be unstable
e.g. against formation of a Wigner-crystal~[\onlinecite{Fr-abel}].

In connection with the $\nu=5/2$ state, it is usually assumed that the cyclotron gap
is quite large and the electrons from the filled lowest Landau level
(with one spin-up and one spin-down electron per orbital) do not participate in the
formation of the strongly correlated state.
Electrons in the {\it half-filled} second Landau level
form a strongly correlated  $\nu=1/2$ incompressible state which we study here.
(An alternative picture for the case of non-chiral states has been considered  in~[\onlinecite{Overbosch}]).

Equation \eqref{charge1} implies that the electron operator
in the effective theory may be written as
\begin{equation}
  \label{electron}
  \psi_{e}(x) = e^{i \sqrt 2 \phi(x) } W(x) .
\end{equation}
where $W(x)$ describes neutral degrees of freedom.
Note that neither $W(x)$ nor $e^{i \sqrt 2 \phi}$ must be
local fields in the field content of the effective CFT. By
Eqs.\,(\ref{J0}--\ref{charge1}) it is seen that  $\psi_{e}(x)$ has unit charge.
The need for additional degrees of
freedom described by $W(x)$ becomes clear if one computes the commutator of two
operators $e^{i\sqrt\phi}:$
\begin{equation}
  \label{perm}
  e^{i\sqrt 2 \phi(x)}  e^{i\sqrt 2 \phi(y)} = e^{-i\theta}
  e^{i\sqrt 2 \phi(y)}e^{i \sqrt 2 \phi(x)}
\end{equation}
where the statistical parameter $\theta=2\pi$ (see
Eq.~\eqref{comphi}). Thus, if $W=1$ the operator $\psi_e$ would have
Bose statistics and hence cannot describe an electron. It is
conceivable that there might exist QHE states violating condition (B).
Such states may exist in systems with an attractive interaction
between electrons forcing them to form strongly bound pairs
[\onlinecite{K8}]. This scenario, although interesting, is unlikely
to be realized in an electronic Hall system and
we do not study it here. Our goal is then
to describe the neutral degrees of freedom of the edge.
We call a chiral CFT ''abelian'' or  ``non-abelian'' depending on whether its primary fields obey abelian or non-abelian statistics, respectively.
In this
letter we limit our analysis to two simple cases: (1) chiral abelian theories
 (2) chiral CFT's
where the neutral sector is decoupled from the charged one and
\begin{equation}
  \label{electron1}
  \psi_{e}(x) = e^{i \sqrt 2 \phi(x) }\otimes W(x) .
\end{equation}
where $W(x)$ is a primary field in a ''non-abelian'' CFT.
In both cases we shall see that there exist plausible theories, that
are in much better agreement with the experiment than, e.g., the
Pfaffian state.

\begin{table}
  \centering
  \begin{tabular}{|c|c|c|c|c|c|c|c|c|c|}
    \hline
   \raisebox{0pt}[0 pt][6 pt]{K}& $(^1_3)$  & $(\begin{subarray}{c} -1 \\ 5 \end{subarray})$
   & $  \left( \begin{subarray}{l} 2, 1;2
        \\  3,3,3 \end{subarray}\right) $ &
    $\left( \begin{subarray}{l} 1, 2;2   \\ 3,3,5 \end{subarray}\right)$ &
    $  \left( \begin{subarray}{l} 1, 1;1  \\  3,5,5 \end{subarray}\right) $ &
    $  \left( \begin{subarray}{l} 4, 0;-1  \\  5,5,5 \end{subarray}\right) $ &
    $  \left( \begin{subarray}{l} 3, 1;-1  \\  5,5,5 \end{subarray}\right) $ &
    $  \left( \begin{subarray}{l} 2, 2;-1  \\  5,5,5 \end{subarray}\right) $
    \\
    \hline \hline \raisebox{0pt}[0 pt][6 pt]{
    $e^*$ } & $\frac{1}{4}$ & $\frac{1}{4}$& $\frac{1}{ 2}$
    & $\frac{1}{4}$ &   $\frac{1}{ 8} $  & $\frac{1}{4}$
    & $\frac{1}{4}$& $\frac{1}{4}$ \\
    \hline
   \raisebox{0pt}[0 pt][6 pt]{ $g$} & $\frac{3}{8}$
   & $\frac{5}{24}$ &  $\frac{1}{2}$  & $\frac{11}{24}$& $\frac{7}{32} $
    & $\frac{9}{40}$&
    $\frac{1}{4}$& $\frac{7}{24}$ \\
    \hline
   \end{tabular}
   \caption{The charges $e^*$ and the scaling dimensions $g$ of the most
     relevant tunneling operators in $N=2,3$ theories. The
     parameters of the $K$-matrices defined in
     Eq.~\eqref{k-gener11} are shown by the symbols $(^b_a)$ for
     $N=2$ and  $(_{l_1, l_2, l_3}^{a_1, a_2;b}),$ for $N=3.$}   \label{tab:3fields}
\end{table}

{\bf Abelian theories.} Chiral abelian edge CFT's are constructed
from a multiplet $\bm \phi=(\phi_1, \dots,\phi_N)$ of free chiral
bosons satisfying $[ \phi_{i}(x), \phi_{j}(x')]= i \pi
\delta_{ij}\mathrm{sgn}(x-x')$ that give rise to $N$ conserved
currents $J^{\mu}_i=(2\pi)^{-1}\epsilon^{\mu\nu}\partial_{\nu} \phi_i.$ The electric current is
a linear combination \mbox{$J_{\rm el}= \mathbf q \cdot \mathbf
J\equiv \sum_i q_i J_i,$} where $q_{i}$ are some coefficients. A
general excitation is described by a vertex operator
\begin{equation}
\psi_{\mathbf v} =e^{i \mathbf{v} \cdot \bm \phi } \label{vo}
\end{equation}
with the statistical parameter $\theta=\pi \mathbf v\cdot \mathbf v$
and the charge $Q_{\rm el}= \mathbf q \cdot \mathbf v.$ If the
operator \eqref{vo} represents an electron, $\theta$ must be odd and
$Q_{\rm el}=1.$ Imposing this condition, we find N solutions $\mathbf
v= \mathbf e_{\alpha}, \; \alpha=1 \dots N.$ The theory is
characterized by its K-matrix $K_{\alpha \mu}=\mathbf
e_\alpha \cdot \mathbf e_\mu,$ whose entries are mutual statistical
phases of electron operators. Using anomaly cancelation condition
\mbox{$\mathbf q\cdot \mathbf q=\nu$} one finds
\begin{equation}
\label{sum} \mathbf Q K^{-1} \mathbf Q =\nu, \qquad \mathbf Q =(1,1,
\dots, 1)
\end{equation}
Condition (C) implies that an arbitrary excitation \eqref{vo}
satisfies $\mathbf v\cdot \mathbf e_\alpha = n_\alpha,$ where
$n_\alpha \in \mathbb Z.$ The conformal spin and the electric charge
of such an excitation are given by
\begin{equation}
h(\mathbf n) = \mathbf n K^{-1} \mathbf n ,  \hspace{10pt} Q_{\rm
el}(\mathbf n)  =
 \mathbf Q  K^{-1} \mathbf n. \label{charg-v}
\end{equation}

Eqs. \eqref{sum}, \eqref{charg-v} can be used for a complete
classification of abelian $\nu=1/2$ states. As an illustration, we
discuss $N=2$ abelian states and
state the results for  $N=3.$ In Ref. [\onlinecite{Fr-abel}] it has been shown that, for physically
interesting states with small relative angular momentum of
electrons, $K$-matrices can always be chosen in the form: {\small
\begin{equation}
\label{k-gener11} K = \left(
 \begin{array}{cc}
   a & b \\
   b & a \\
 \end{array}
\right),\qquad
 K=\left(\begin{matrix}
     l_{1}& a_1& a_2\\
     a_1 & l_2 & b\\
     a_2 & b & l_{3}
   \end{matrix}\right),
\end{equation}
} where $l_{1}\le l_2\le l_{3}$.
For $N=2$ fields, (\ref{sum})
reduces to
\begin{equation}
  \nu =
  \frac{2}{a+b}, \hspace{10pt} a+b =4,\hspace{10pt}a\in 2\mathbb{Z}+1 \hspace{10pt}  ( \nu = 1/2)
\label{ab}
\end{equation}
In such a theory the electric charge is given by : $Q_{\rm em} =
(n_1+n_2)/(a+b)= (n_1+n_2)/4$.  There are only three $K$-matrices (with $\det K>0$, for the theory to be chiral) for which the
conformal spin of an electron $h_{e}\leq7/2.$

A complete classification of irreducible $N=3$ theories can be found in
Ref.~[\onlinecite{Fr-abel}]. There are six distinct indecomposable
three-dimensional \emph{chiral} lattices describing admissible $\nu=1/2$ QHE
states. The parameters of the $K$-matrices, the charges and scaling dimensions
of the most relevant tunneling operators in all these theories (calculated
using the Eq.~(\ref{charg-v})) are given in Table~\ref{tab:3fields}.
Comparing with experiment (see Fig.~1(a)), one can see that there are two good
candidate theories, both with $N=3.$ One of them has minimal charge
$e^{*}={1}/{8}$.

{\bf Non-abelian theories.}
A generalization of the above construction may be obtained
if one
replaces the Heisenberg algebra describing
the modes of free bosons by a central extension of some
Lie algebra $\mathfrak g.$ The resulting theory contains several
non-abelian currents $J_a$ satisfying the Kac-Moody (KM) algebra
$\widehat{\mathfrak g}_{k}:$
\begin{equation}
[J_{a}(x), J_{b}(0)]=2 \pi i k \delta_{ab} \delta'(x)+ 2\pi i \delta(x)
\sum_{c}f_{ab}^c J_c (0),
\label{KM}
\end{equation}
where $f_{ab}^c$ are the structure constants of $\mathfrak g$ and
$k$ is a parameter called  the {\it level}. We do not attempt to
explore all such theories. Instead, we focus our attention on theories
where (a) the electric current commutes with other KM currents; (b)
the electron operator is given by Eq.~\eqref{electron1}, where $W$
is a KM primary field; (c)  $\mathfrak g$ is a simple affine Lie
algebra. In the following we call the corresponding CFT the {\it
neutral sector}. All non-abelian $\nu=5/2$
states poposed so far are of this type. Unitary CFT's associated with Lie algebras
are Wess-Zumino-Witten (WZW) models or coset theories generated from WZW models by means of the so-called GKO construction (see e.g. [\onlinecite{Difrancesco}]). Among them are theories
based on KM algebras at level 1, which may give rise to ``abelian'' CFT's, see~[\onlinecite{Fr-abel, Fr-non-abel}]
for applications to the QHE~\footnote{In fact, the theory at level 1 is abelian only if
$\mathfrak g_{k}$ belongs to A,D,E series}.

The requirement that the electron operator has Fermi
statistics imposes $W(x) W(y) = - W(y) W(x),$ i.e. the neutral
sector must contain a primary field of half-integer conformal spin.
Not every such field is, however, acceptable. Indeed, the OPE of a
pair of primary fields in the neutral sector is generally given by
\begin{equation}
\phi_a(z)\phi_b(0)=\sum_{c} C_{ab}^c \phi_c z^{h_c-h_a-h_b} +
\text{descendants},
\label{OPE}
\end{equation}
$h_i$ is the conformal dimension of the field $\phi_i.$ In general,
the dimensions $h_i$ are not commensurate. Substituting $W$ instead
of $\phi_a$ in Eq.~\eqref{OPE} one can see that, in order to satisfy
the locality requirement (C), the right hand side of Eq.~\eqref{OPE}
must contain exactly one primary field for every $\phi_b.$ In the
language of fusion rules this is expressed as $W\times \phi_b =
\phi_c,$ i.e. fusion with  $W$ determines a permutation on the set of
primary fields. Such a primary field $W$ is called a {\it simple
current} [\onlinecite{Schellekens}]. The presence of a half-integer-spin simple current in the
theory is a strong constraint.

Quasihole excitations are described by
operators of the neutral and the charged sectors~as
\begin{equation}
\psi_{\rm qh}=e^{i q \sqrt 2 \phi(x) }\otimes V(x)
\label{qh}
\end{equation}
where $q$ is the quasihole charge and $V(x)$ is a Virasoro-primary field
in the neutral sector.
Substituting Eqs. \eqref{qh}
and \eqref{electron1} in \eqref{OPE} and imposing the locality constraint
one finds
\begin{equation}
h_{W\times V}-h_W-h_V+2 q\in \mathbb Z
\label{fuscharge}
\end{equation}

An important property of spectrum of electric charges of the edge CFT is
expressed in terms of the {\it order} of the simple current $W$,
defined~\cite{Schellekens} as the smallest integer $\ell$ such that
$W^{\ell}=\mathbb{I}$.  It is shown~[\onlinecite{Fr-non-abel}] that
\begin{equation}
q \in \frac{1}{\ell d_{H}}\mathbb Z, \label{charges}
\end{equation}
where $d_{H}$ is the \emph{Hall denominator}, $d_{H}=2$ in our case.
The dimension of the tunneling operator $\psi_{\rm qh}^\dagger \psi_{\rm qh}$
is
\begin{equation}
g=2( h_V+q^2).
\label{gofV}
\end{equation}

{\bf Wess-Zumino-Witten models.}
One of the early proposals for
a non-abelian neutral sector is the $\widehat{\mathrm{su}}(2)_2$ theory~\cite{WenSU22}.
This theory has
central charge $c=3/2$ and contains an $\rm SU(2)$ triplet of  Majorana-Weyl
fermions with $h_W=1/2$ (order $\ell=2$
simple currents) and a doublet of quasihole
excitations with $h_{\rm qh}=3/16.$ In this theory
$W$ is identified with the Majorana-Weyl triplet. The charge of the most relevant
quasihole is $e^*=1/4$ and  the  dimension of the
tunneling operator is $g=2(3/16+1/16)=1/2.$ Among the theories discussed in [\onlinecite{Radu2008}]
this model fits the experiment data best.

To generalize the  $\widehat{\rm su}(2)_2$ theory, one may consider
$\widehat{\rm su}(2)$ at higher levels or different Lie algebras. With
increasing level $k$ or rank $r$ of $\mathfrak g$, the
complexity of the CFT increases, while the smallest dimension $h_V$
decreases. From these observations one can deduce that
only a limited number of WZW models are plausible
candidates compatible with experiment data.  This allows one
to obtain all plausible WZW theories on a case by case basis.

The interesting candidate theories obtains as a result of this analysis are
described in Table II.
\begin{table}
  \begin{tabular}{|c|c|c|c|c|c|}
    \hline
    Model & $\widehat{\mathrm{su}}(2)_6,$ & $\widehat{\mathrm{sp}}(4)_1$ & $\widehat{\mathrm{sp}}(4)_3$
    & $ \rm \widehat {so}(7)_1 /\widehat{ su}(2)_{1}$ & $ (\hat G_2)_2 /\widehat{\rm su}(2)_{2}$  \\
    \hline
    $c$ & 9/4 & $5/2$ & $5$ & $5/2$ & $19/6$\\
    \hline
    $P$ & 7 & $3$ & $ 10$ & $6$ & $12$ \\
    \hline
    $h_e$ & 5/2 & $3/2$ & $5/2$ & $3/2$ & $3/2$\\
    \hline
    $g$ & 5/16 & $3/4$ & $13/24$ & $1/2$ & $5/12$\\
    \hline
   \end{tabular}
   \caption{ The central charge $c,$ the number $P$ of KM
   primary fields,
  the conformal spin $h_e$ of the electron
   operator  and the tunneling dimension $g$ for  non-abelian  models
   ($\nu=1/2$).}
 \label{tab:WZW}
\end{table}
The smallest fractional charge $e^*$ in all these theories is $1/4.$
Comparison with experiment is shown in Fig.~1(c). The model based on
$\widehat{\mathrm{sp}}(4)_1$ is the simplest one and is similar to
$\widehat{\mathrm{su}}(2)_2$ containing one multiplet of electrons and one of
quasiholes. It, however, predicts too large a value of $g.$ The best fit to
the experiment corresponds to $\widehat{\mathrm{sp}}(4)_3.$

\begin{widetext}
\begin{figure*}
\includegraphics[width=\textwidth]{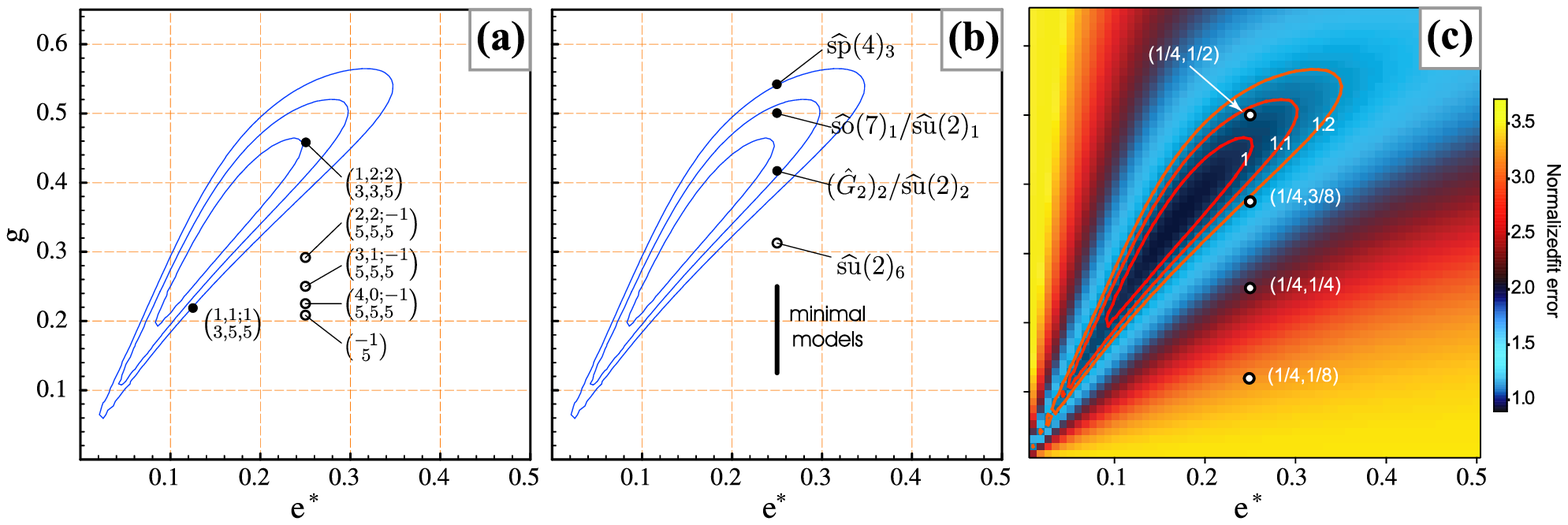}
\caption{Predictions of abelian, (a), and non-abelian, (b), theories
discussed in the present work are shown on the $(e^*, g)$ plane and
compared with the fit quality map, (c), from
Ref.~\onlinecite{Radu2008}. Theories lying within the $3\sigma$
contour are shown as full circles. Points corresponding to
previously discussed proposals are shown in (c). In particular, the
point $(1/4, 1/2)$ in (c) corresponds to the $\widehat{\rm su}(2)_2$
and the antipfaffian states and $(1/4, 1/4)$ to the Moore-Read
state. }
  \label{fig:exp}
\end{figure*}
\end{widetext}

{\bf Coset models.} A much bigger class of conformal field theories
(all known rational unitary CFT's) is obtained via the GKO coset
construction. For example, the Moore-Read state~\cite{Moore-Read} is
associated with the coset $ \rm \widehat{ su}(2)_1 \oplus \widehat{
su}(2)_1/\widehat{ su}(2)_{2}$ describing the chiral sector of the
critical Ising model. An exhaustive analysis of the possibilities
offered by the coset construction is the subject of future work. Here
we discuss some simple examples.

{\it (a) Virasoro minimal models.} These form an infinite series of CFT's
associated with the cosets $ \rm \widehat {su}(2)_k \oplus \widehat{
  su}(2)_1/\widehat{ su}(2)_{k+1}.$ The properties of these models, which
exhaust all unitary CFT's with central charge $c<1$, are described in
literature in great detail.  For $k=4 m-3$ and $k=4m-2,$ $m\in \mathbb Z$ the
model contains an order $2$ fermion simple current, identified with $W.$ The
conformal spin of this current grows rapidly with $m,$ and becomes $15/2$
already for $m=2,$ so that theories with $m>2$ are implausible from the point
of view of stability.  For all models in the series one finds the minimal
charge $e^*=1/4$ and the dimension $g$ of the most relevant tunneling operator
is monotonically decreasing from $g=1/4$ to $g=1/8$ (a solid line in
Fig.1(b)).  Thus, in this series the Moore-Read state gives the best fit to
the experiment and is preferred from the point of view of stability.

(b) Another interesting class of coset models with the central
charges $c\geq 1$ are \emph{super-Virasoro minimal models.} For our
analysis, supersymmetry just means that a simple current with
conformal weight $3/2$ is present in the spectrum. In this series $e^*=1/4$.
The dimensions $g$ of the most relevant tunneling
operators lie between $g=59/280$ and $g=1/8$ (solid line in
the Fig.~1(b)).

(c) As examples of more general coset models we mention here  $ \rm \widehat
{so}(7)_1 /\widehat{ su}(2)_{1}$ and   $ (\hat G_2)_2 /\widehat{\rm
su}(2)_{2}$ (see Fig.~1(b) and Table II).

In conclusion, combining the fundamental requirements (A-C) with bounds
imposed by the experiment [\onlinecite{Radu2008}] we have shown that there is
a limited number of chiral conformal field theories that may serve as
effective field theories for the $\nu=5/2$ edge (see Fig.~1 (b)-(c)). Some of these
theories that have not been previously discussed.
Intriguingly, there exist both abelian and non-abelian states with exactly the
same values for $e^*$ and $g$ as the Pfaffian, anti-Pfaffian and $\widehat{\rm
  su}(2)_2$ states used in [\onlinecite{Radu2008}] for comparison. Thus, it is
impossible to distinguish between these states and, in particular, reveal
their non-abelian nature based on the tunneling data of
[\onlinecite{Radu2008}] \emph{only}.  The ambiguity might be resolved with the
help of further experimental data. In particular, information on the zero-bias
cusp in the tunneling density of states of electrons~[\onlinecite{Chang}]
would set bounds on the maximal value of the conformal spin of the
electron; noise measurements or coulomb-blockade experiments might be used to
verify the theory predicting minimal charge $1/8;$ interferometric experiments
using Mach-Zehnder geometry might be used to detect conformal dimensions of
different excitations as proposed in Ref.~[\onlinecite{Levkivskyi:08}].

We thank I. Radu for providing experimental data and O. Ruchayskiy  and 
R. Morf for valuable discussions.

\end{document}